
%
%
  \MAINTITLE{
Abundance analysis and origin of the $\zeta$\,Sculptoris open cluster\FOOTNOTE{
Based on observations carried out at the European Southern Observatory,
La Silla, Chile}
\FOOTNOTE {Table~3 is only available in electronic form: see the editorial in
A\&A 1992, Vol. 266 No 2, page E1}
}
  \AUTHOR{ B. Edvardsson, B. Pettersson, M. Kharrazi, B. Westerlund }
  \INSTITUTE={
Uppsala Astronomical Observatory, Box 515, S-751 20 ~Uppsala, Sweden}
  \ABSTRACT={
We have determined chemical abundances and
radial velocities for stars in the field of the $\zeta$\,Sculptoris cluster.
We find that the cluster metal deficiency previously found from $UBV$
photometry
is too high; the cluster overall metallicity, [Fe/H]$\;=+0.24$, is about 70\%
higher than the solar value.
The chemical abundance pattern is unusual: the Ni/Fe ratio is significantly
lower than typical for field stars with similar Fe and Ca abundances, and
the cluster may also be deficient in the $\alpha$ element Si.
For its age, the cluster is unusually far from the galactic disk,
$\approx$240\,pc.
The adopted heliocentric cluster radial velocity, +3.9 km\,s$^{-1}$, shows that
it is close to its maximum distance from the galactic plane.
We suggest that the $\zeta$\,Sculptoris cluster was formed in the galactic disk
45\,Myr ago by the interaction of a high velocity cloud with the interstellar
medium, and that its formation may be connected with that of Gould's Belt.

}
  \KEYWORDS={
open clusters and associations: individual: $\zeta$\,Scl --
open clusters and associations: individual: Gould's Belt --
Galaxy: abundances --
Stars: abundances
}
  \THESAURUS={5 (10.15.2 $\zeta$ Scl; 10.15.2 Gould's Belt; 10.01.1; 08.01.1)
 }
  \OFFPRINTS={ Bertil Pettersson }      
  \DATE={ Received December 20, 1993, accepted May ??, 1994 }
  \maketitle
 \MAINTITLERUNNINGHEAD{The $\zeta$\,Sculptoris cluster}
 \AUTHORRUNNINGHEAD{Edvardsson et~al.}
\titlea{Introduction}
The $\zeta$ Sculptoris cluster, also known as Blanco\,1, is about 240\ts pc
from
the Sun and at that distance from the galactic plane; its galactic latitude is
$-79$\degr.  It is a relatively young cluster.  Its age is about 50 Myr
(Perry et~al. 1978, de Epstein and Epstein 1985).  A few years ago some of us
(BP \& BW) carried out a
photometric study (in $UBV$ and $uvby\beta$) of 130 stars in the cluster region
(Westerlund et~al. 1988, henceforth WGLPB) and determined cluster membership,
interstellar reddening ($E_{B-V}$\,=\,0\fm02), distances to the stars and
metal content.  The majority of the field F stars were found to have $-0.2 \leq
[\rm{Me/H}] \leq 0$; for the most metal deficient star [Me/H]\,=\,$-0.9$.  The
cluster was found to be slightly metal deficient with [Me/H] $\leq -0.17$ and
probably somewhat more deficient than the field F stars at the same distance
(see WGLPB, Fig.\ 13).  A young cluster with these properties is of great
interest in particular as it is rather far from the galactic plane.

The mean [Me/H] given above is, however, somewhat uncertain.
The individual values scatter appreciably, the dispersion is $\sigma =
\pm\ts0.11$. The $\delta\rm{m_0}$ values, derived from the $uvby$ photometry of
8 stars, gave [Me/H]  = $-0.08$, so that the mean value above, based
on 19 stars, depend strongly on the determinations with the aid of the
$\delta(U-B)_0$ values. We have therefore found it desirable to determine
the metallicity spectroscopically using available techniques to obtain chemical
abundances. Here, we present the results of a spectroscopic analysis of 9
stars, 5 of which were suggested to be cluster members by WGLPB.

We have also obtained CORAVEL radial velocities for 12 stars in the cluster
region and expect to obtain proper motion data from HIPPARCOS observations.  A
redetermination of the space motion of the cluster will then be possible.
\newdimen\digitwidth
\setbox0=\hbox{\rm0}
\digitwidth=\wd0
\begtabfull
\tabcap{1} {Observational data for the HR stars and a comparison with other
work, see text for details.  Velocity data and quality, Q, for three stars in
common with the Mt Wilson catalogue of radial velocities (Wilson, 1953) are
given in columns 6 and 7 and data communicated by J.~Andersen are given in
columns 8 and 9.  The radial
velocities of this paper, $RV$, have been adjusted to the system of Mt Wilson
by
adding 0.6 km\,s$^{-1}$
}
\vbox{
\catcode`~=\active
\def~{\kern\digitwidth}
\def\:{\kern\digitwidth}
{\halign {
{}~#\hfil&\hskip1.7mm\hfil#\hfil&\hskip1.7mm\hfil#\hfil&\hskip1.7mm\hfil#\hfil&
\hskip1.7mm\hfil#\hfil&\hskip1.7mm\hfil#\hfil&\hskip1.7mm\hfil#\hfil&
\hskip1.7mm\hfil#\hfil&\hskip1.7mm\hfil#\hfil\cr
\multispan9\hrulefill\hfil\cr
Star&$RV$&$\overline\epsilon$&n&FWHM&$RV_{\rm MW}$&Q&$RV_{\rm JA}$&
$\overline\epsilon$\cr
 No.\hfil&[km/s]&    & & [\AA] &[km/s]&&[km/s]&\cr
\multispan9\hrulefill\hfil\cr
HR\,33   &  ~14.8&~.1 & ~77& 0.28 &   ~14.8& A &   &    \cr
HR\,7126 &$-$45.0&~.2 & ~54& 0.52 & $-41.8$& B &   &    \cr
HR\,7232 &  ~60.0&~.1 & 116& 0.25 & \omit&\omit  &59.6&.1  \cr
HR\,8181 &$-30.2$&~.2 & ~90& 0.28 & $-30.2$& A &   &    \cr
\multispan9\hrulefill\hfil\cr
}}}
\endtab

\titlea {Observations and data reductions}
\titleb {Selection of the observed sample}
The stars to be observed were selected to meet three basic criteria -- (i) most
stars should be probable members of the cluster, (ii) a few field stars at the
cluster distance should be included and (iii) the spectral type
should be later than F0.
The cluster memberships of the proposed stars were assessed
both from their photometric distances and their positions in the colour-colour
diagrams of WGLPB (their member stars are marked by asterisks in Table~III of
WGLPB).  The need for a particular range of spectral types was dictated by the
method of abundance analysis described in Sect.\ 3.2.  CoD\,$-$29\degr\,18906
(No.\ 5 in WGLPB) was included because the
photoelectric $uvby\beta$ data indicated that it has a significantly lower
metal
content than any other star in the cluster region.  Although this suggested a
non-membership it was nevertheless regarded as interesting enough to qualify
for
observing.  17 stars were selected for observing, 12 of them were actually
observed.  To enable a comparison with other work, a number of bright stars
from
a paper by Edvardsson et~al. (1993a, EAGLNT hereafter) were also observed, see
Tables~1 and 2 for details about these and about the cluster sample.
\begtabfullwid
\tabcap{2} {Observational data for the stars.  Apart from the self-evident
columns, col.\ 6, $\overline\epsilon$, gives the internal mean error in the
CASPEC radial
velocities, col.\ 7, n, the number of spectral lines used to determine each
velocity, col.\ 10, $\overline\epsilon$, the error in Lindgren's velocities
($RV_{\rm HL}$) and col.\ 12, $\overline\sigma$, the error of the mean for the
Mermilliod-Mayor sample ($RV_{\rm MM}$).
The $RV_{\rm MM}$ velocities are mean values of between 2 and 5 separate
measurements.
Finally, the straight mean radial
velocities from columns 5, 9 and 11 and their standard deviations are given in
columns 13 and 14
}
\vbox{
\catcode`~=\active
\def~{\kern\digitwidth}
\def\:{\kern\digitwidth}
{\halign {
{}~#\hfil&\hfil#\hfil&\quad\hfil#\hfil&\quad\hfil#\hfil&\quad\hfil#\hfil&
\quad\hfil#\hfil&\quad\hfil#\hfil&\quad\hfil#\hfil&
\quad\hfil#\hfil&\quad\hfil#\hfil&\quad\hfil#\hfil&\quad\hfil#\hfil&
\quad\hfil#\hfil&\quad\hfil#\hfil\cr
\multispan{14}\hrulefill\hfil\cr
Star \hfil&CoD&V$^1$&$V_{\rm o}-M_V^1$&$RV$&$\overline\epsilon$&n&FWHM&
$RV_{\rm HL}$&$\overline\epsilon$&$RV_{\rm MM}$&$\overline\sigma$&
$\overline{RV}$&$\overline\sigma$\cr
 No.$^1$\hfil    &  &      &        &[km\,s$^{-1}$]&    & & [\AA] &
[km\,s$^{-1}$]&&[km\,s$^{-1}$]&&[km\,s$^{-1}$]&\cr
\multispan{14}\hrulefill\hfil\cr
  5          & $-29\degr18906$ &10.70&6.42&   ~53.7&0.3 & ~24 & 0.27 &        &
   &    && &\cr
  8          & $-29\degr18909$ &10.10&6.58&   ~~3.8&0.2 & ~51 & 0.48 &   ~~4.6&
{}~.5&~4.9&0.3&~~4.4&0.6\cr
 28*         & $-31\degr19554$ &10.51&6.83&   ~~4.8&0.1 & ~69 & 0.40 &   ~~4.8&
{}~.4&~5.2&0.6&~~4.9&0.2\cr
 38*         & $-30\degr19787$ &10.72&6.91&   ~~6.5&1.1 & ~23 & 1.61 &        &
   &    &&\cr
 47          & $-29\degr18938$ &10.23&6.90&   ~~0.3&0.4 & ~29 & 1.03 &   ~~0.8&
2.0&~6.0&3.6&~~2.4&3.2\cr
 54*         & $-31\degr19572$ &10.35&7.04& ~$-4.1$&0.8 & ~30 & 1.44 & ~$-7.0$&
1   &    &&~$-5.6$&2.0\cr
 57*         & $-30\degr19801$ &10.38&7.06& $-15.6$&0.4 & ~54 & 0.73 & $-26.0$&
1.0&    &&$-20.8$&7.4\cr
 60*         & $-31\degr19574$ &10.61&6.91&   ~~5.1&0.3 & ~43 & 0.65 &   ~~5.1&
{}~.8&~5.2&0.5&~~5.1&0.1\cr
 63*         & $-31\degr19576$ &10.60&6.93&   ~~3.8&0.1 & ~79 & 0.32 &   ~~3.7&
{}~.5&~3.7&0.6&~~3.7&0.1\cr
 70*         & $-31\degr19579$ &11.12&6.92&   ~~3.2&0.2 & ~68 & 0.53 &   ~~8.1&
{}~.7&~4.6&1.6&~~5.3&2.5\cr
121*         & $-31\degr6~~~~$ &10.22&6.64&   ~12.2&2.6 & ~~8 & 2.76 &        &
   &    &&&\cr
125          & $-30\degr6~~~~$ &10.59&6.94&   ~13.2&0.3 & ~49 & 0.70 &   ~~1.7&
1.8&$-2.2$&1.9&~~4.2&8.0\cr
\multispan{14}\hrulefill\hfil\cr
\multispan{14} $^1$ From Westerlund et~al.\ (1988) (WGLPB).\hfil\cr
\multispan{14} * Cluster member according to WGLPB\hfil\cr
}}}
\endtab
\titleb {Observations and reductions}
The stars were observed with the ESO 3.6\ts m telescope at La Silla
in Chile during August 20--24, 1991.  We used the CASPEC echelle spectrograph
operated at a spectral resolving power of 28\,000 in the spectral range
5450\ts\AA --6170\ts\AA.  The detector was a Tektronix 512x512 pixels CCD chip,
characterized by a low read-out noise ($\sim$8 e$^-$) and a negligible dark
current.  All spectra were exposed to an S/N per pixel of between 100 and 400.
The data was reduced with the MIDAS package from ESO.  The spectra were
wavelength-calibrated against a Thorium spectral lamp and rebinned to a linear
spectral resolution of 0.085\ts\AA\ per bin, closely corresponding to the
original pixel size.  The typical error (standard deviation) in the calibration
is 0.013\ts\AA, corresponding to an uncertainty in the radial velocity of a
given spectral line of $\pm$0.65 km\ts s$^{-1}$.  The separate spectral orders
were then extracted, rectified and normalized to an arbitrary continuum level
of
unity.  The continuum normalization was made by fitting a cubic spline curve
through a number of continuum windows in each echelle order, selected from
the Solar Flux Atlas by Kurucz et~al.\ (1984).  The mean FWHM of the
instrumental profile as measured from the night sky emission lines was
0.20\ts\AA$\pm$0.01\ts\AA.
\begfig 6.0 cm
\figure{1} {A plot of one of the spectral orders from two stars; HR\,7126, a
bright comparison star, and WGLPB No.\ 70, an 11:th magnitude cluster member.
The spectra are normalized to a continuum level of 1.0 and the tracing of No.\
70 was shifted by $-$0.15 to facilitate comparison.  The spectral absorption
lines used in this spectral order for the chemical abundance
analysis are indicated}
\endfig
\titleb {Measurement of equivalent widths, W$_\lambda$}
The equivalent widths of the photospheric absorption lines in the spectra
were interactively measured on a computer by fitting a single or double
Gaussian
function to each identified line profile within a narrow spectral window.
Reliable equivalent widths in stars Nos.~38, 54 and 121 could not be obtained
due to line blending caused by large rotational line broadening in these young
 stars.
They have therefore been excluded from the ensuing analysis.
The equivalent widths measured are generally between 20 and 120\,m\AA,
sometimes
larger, which makes the abundance results sensitive to the assumed
microturbulence parameters.
The random errors in the measurements are estimated to be typically 4--5\,m\AA.
A few of the measured lines were excluded from the final analysis since the
resulting abundances from those lines in several stars were found to deviate
systematically (as a function of effective temperature) from those of other
lines of the same species.
The causes of these deviations are believed to be unknown line blends or line
misidentifications.
The measured equivalent widths are given in Table~3 (available in
electronic form).
One of the spectral orders is plotted in Fig.~1 for both
a cluster star and a bright comparison star.
\titleb {Radial velocities}
The radial velocities for the stars were determined by fitting Gaussian curves
to the absorption lines in the extracted and normalized spectra.  As the
rotational broadening is unimportant for determining the line centres all 16
observed stars could be measured.  The final radial velocity, $RV$, for each
star
was obtained by averaging all measured lines within 2 standard deviations of
the
resulting mean.  The number of lines finally used for each star is given in
column 4 of Table~1 for the comparison stars and in column 7 of Table~2 for the
programme stars.  The observed velocities were then corrected for the orbital
motion of the Earth and for the rotation around its axis to obtain the
heliocentric radial velocities.  The final results, given in column 2 of
Table~1
and column 5 of Table~2 were obtained by adding 0.6 km\,s$^{-1}$ according to
the the discussion in Sect. 3.1.
The mean error, $\overline\epsilon$, for each
star is also given in column 3 in Table~1 and column 6 in Table~2.

\titlea {Analysis and results}
\titleb {Radial velocities}
In order to check the quality of our CASPEC radial velocities we compared our
data set to data obtained both from literature searches and from
private communication.

The data for the four HR-stars were compared to measurements found in the Mt
Wilson general catalogue of stellar radial velocities (Wilson 1953, MW below)
for HR\,33, 7126 and 8181, and to CORAVEL data obtained from J.\ Andersen
(private communication) in the case of HR\,7232 (see Table~1 for details).
For HR\,33 and HR\,8181 we found a small difference between our data and those
of MW of 0.6 km\,s$^{-1}$, with MW's data being more positive.  For
HR\,7126 the difference is larger. However, this star is a suspected radial
velocity variable according to the Bright Star catalogue; thus, we will not
consider it in our discussion.  For HR\,7232 no entry could be found in MW,
but we have a velocity communicated by J.\ Andersen who obtained 59.6
km\,s$^{-1}$ with the CORAVEL.
If we consider only the two stars in common with MW that have stable
velocities,
we arrive at a zero point shift $RV_{\rm MW}$ -RV = 0.6\,km\,s$^{-1}$.  If we
include HR\,7232 we arrive at a shift 0.4 km\,s$^{-1}$.  We prefer, however, to
consider only the MW data and adopt a zero-point correction of +0.6
km\,s$^{-1}$
in the following.

Several of the  stars were observed with the CORAVEL instrument on the
Danish 1.5\,m telescope at ESO, both, kindly at our request, by H.\ Lindgren
(HL
in the following) and, independently, by J.C.\ Mermilliod and M.\ Mayor (1993,
MM).  7 of the stars, Nos.\ 8, 28, 47, 60, 63, 70 and 125 were observed by all
of us, thus, we have three independent sets of measurements for these stars
(see Table~2).
HL also observed stars 54 and 57.  After applying the
correction of +0.6 km\,s$^{-1}$ to our CASPEC data it appears that the CASPEC
and CORAVEL data sets are generally very well correlated.  Especially it brings
the CASPEC data in accordance with the CORAVEL data for the stars with a small
($\leq$0.6 km\,s$^{-1}$) velocity spread, like Nos.\ 28, 60 and 63.  We can
obviously order the stars into two groups, one group containing the stars with
a small external velocity spread, stars No.\ 8, 28, 60 and 63 and another group
with stars with more deviating velocities Nos.\ 47, 54, 57, 70 and 125.
Two stars, Nos.\ 38 and 121, have line widths too large for the CORAVEL to
measure and they and No.\ 5 were only measured using CASPEC data.

Let us comment briefly on the two groups.  The 4 stars with a small spread
between the different data sets have mean velocities very similar
to one another. It seems likely that they share a common motion,
representative of the cluster mean velocity.  Star number 8 was not considered
by WGLPB as a cluster member due to its deviating position in the
$V/(B-V)$-diagram.  However, since it has a velocity very close to the mean of
remaining three, it may still be a member. If we assume that this star is a
binary, but seen almost pole-on, we can account both for the smaller
distance modulus and for the apparently stable radial velocity.

In the other group of five stars it seems clear that the spread in velocities
is
too large to be attributable to observational errors, possibly excluding No.\
54.  We are then led to believe that they are radial velocity variables and as
such likely to be binary stars.  Their mean velocities are not well determined
as only a few points on their velocity curves are measured.  Stars 47, 70 and
125 have three points determined and mean velocities reasonably close to that
of
group one.  These three stars are judged to be binary members of the cluster,
cf.\ also the Discussion in Sect.~5.1.  As for stars 54 and 57, with only two
points on their velocity curves, they may not be considered as cluster members
based on their radial velocity data alone.
\titleb {Chemical abundances}
The chemical analysis was performed with the same programs and methods as were
used in EAGLNT.
A brief description of the separate steps will be given below.
\titlec {Model atmospheres and atmospheric parameters}
The atmospheric models are line-blanketed, plane-parallel, flux-constant and
calculated in the LTE approximation.
The calculations include the new, very extensive, ab initio calculations of
atomic line data by Kurucz (1989), which is shown by EAGLNT to dramatically
improve (compared to earlier theoretical models) the emergent surface fluxes
for
the Sun and other Pop~I stars.

The programme star atmospheric model parameters, effective temperature;
$T_{\rm eff}$, logarithmic surface gravity; $\log~g$ and overall metallicity;
[Me/H], were calculated from the Str\"omgren $uvby$ photometry as described in
EAGLNT.
For this, the reddening of this high galactic latitude cluster was taken to be
$E_{b-y}=0.017$,
which is the mean of $E_{b-y}=0.019$ and $E_{B-V}=0.02$
(corresponding to $E_{b-y}=0.014$)
given in WGLPB.

Also the recipe for the microturbulence parameters, $\xi_{\rm t}$, was
originally adopted from EAGLNT.
These values of $\xi_{\rm t}$ were, however, in course of the analysis found to
be too low, and values 0.5\,km\,s$^{-1}$ higher were chosen for the final
abundance analysis of the cluster stars, see Sect.~4.1.2.

The photometric data, its sources and the model parameters are presented in
Table~4.
\begtabfull
\tabcap {4} {
Str\"omgren $uvby$ photometry for the programme stars analysed,
with references and resulting atmospheric parameters.
The parameters for the four bright comparison stars are also given
}
\vbox {\halign {
#\hfil&\hfil#~\hfil&\hfil#~\hfil&\hfil#~\hfil&\hfil#~\hfil&\hfil#~\hfil&
\hfil#~\hfil&\hfil#~ &\hfil#~\hfil\cr
\multispan{9}\hrulefill\hfil\cr
Star & $b-y$ & $m_1$ & $c_1$ & Ref. & $T_{\rm eff}$ & $\log\,g$ & [Me/H] &
 $\xi_{\rm t}$~ \cr
No &       &       &       &      &      [K]      &        &      &[km/s]\cr
\multispan{9}\hrulefill\hfil\cr
  5     & .311 & .087 & .279 & 1 & 6350 & 4.60 & $-$1.00\rlap{*}~ & 1.4~ \cr
  8     & .287 & .168 & .480 & 1 & 6765 & 4.35 &   +0.00\rlap{*}~ & 2.6~ \cr
 28     & .280 & .152 & .442 & 2 & 6775 & 4.55 & $-$0.20\rlap{*}~ & 2.3~ \cr
 47\rlap{$\dagger$} & (.267 & .110 & .525) & 1 & 7000 & 4.50 &
  +0.00\rlap{*}~ & 2.5~ \cr
 57     & .268 & .156 & .475 & 3 & 6875 & 4.60 & $-$0.20\rlap{*}~ & 1.8~ \cr
 60     & .295 & .140 & .457 & 3 & 6600 & 4.30 & $-$0.35\rlap{*}~ & 2.5~ \cr
 63     & .280 & .180 & .387 & 2 & 6900 & 4.65 &   +0.20\rlap{*}~ & 2.3~ \cr
 70     & .331 & .190 & .335 & 3 & 6470 & 4.55 &   +0.10\rlap{*}~ & 2.1~ \cr
125\rlap{$\dagger$}      & & & & & 6800 & 4.50 &   +0.00\rlap{*}~ & 2.4~ \cr
HR\,33\rlap{$\ddagger$}  & & & & & 6202 & 4.06 & $-$0.52~ & 2.0~ \cr
HR\,7126\rlap{$\ddagger$}& & & & & 6651 & 4.16 & $-$0.01~ & 2.2~ \cr
HR\,7232\rlap{$\ddagger$}& & & & & 5625 & 4.21 &   +0.05~ & 1.3~ \cr
HR\,8181\rlap{$\ddagger$}& & & & & 6143 & 4.34 & $-$0.72~ & 1.6~ \cr
\multispan{9}\hrulefill\hfil\cr
\multispan{9} * Preliminary photometric metallicities \hfill\cr
\multispan{9} $\dagger$ See text for derivation of atmospheric parameters
\hfill\cr
\multispan{9} $\ddagger$ Atmospheric parameters from Edvardsson et~al.\ (1993a)
\hfill\cr
\multispan{9} Ref.\ 1: Westerlund et~al.\ (1988) \hfill\cr
\multispan{9} Ref.\ 2: Eggen (1972) \hfill\cr
\multispan{9} Ref.\ 3: Epstein (1968) \hfill\cr
}}
\endtab

Since we have not been able to find any $uvby$ photometry for No.~125, we
estimated its $T_{\rm eff}$ from comparison with the other programme stars in
the $UBV$ two-colour diagram.
We also reject the $uvby$ data for No.~47 on grounds of the very large
difference between the metallicity derived from a preliminary spectroscopic
analysis and the metallicity derived from the $uvby$ calibration (there is only
one observation of $uvby$).  We draw support for this rejection also from the
large difference between the two [Me/H] calibrations for the star in WGLPB
(Table~IX).  As for No.~125 we estimate its $T_{\rm eff}$ from comparison with
the other programme stars in the $UBV$ two-colour diagram.

If we, alternatively, apply the $\beta - T_{\rm eff}$-calibration of
Saxner \& Hammarb\"ack (1985) and according to EAGLNT add a correction of
+150\,K we would for No.~47 find a 50\,K hotter $T_{\rm eff}$ and for No.~125
a 50\,K cooler $T_{\rm eff}$ than the values adopted in Table~4.

The surface gravity and metallicity for No.~47 and No.~125 were given typical
values for the analysed stars, and the microturbulence parameter calculated
like
for the others.

In the final analysis, new models with metallicities taken from a preliminary
abundance analysis were used, but the effects of this change were found to be
very small.
\titlec {Chemical abundances and line data}
The chemical abundances were derived from the equivalent widths measured in
Sect.~2.3.  Using an atmospheric model and available line data for each
spectral
line, a FORTRAN program was used to calculate the surface fluxes in a number of
wavelength points in the spectral line as well as the continuum flux for the
vicinity of the line from the appropriate model atmosphere.
The program further integrated the flux removed by the spectral line to
calculate a theoretical equivalent width, and compared it to the width measured
in the observed spectrum.  By automatically iterating the calculation with
improved estimates for the abundance of the chemical element until consistency
was achieved, a measure of the abundance of the line-forming element in the
star
was derived.
When a species was represented by more than one line in a stellar spectrum,
a mean value and the standard deviation of the mean was also calculated.

\begtabfullwid
\tabcap{5} {
Atomic line data and Solar equivalent widths.
The columns give, respectively, the line rest wavelength, the excitation energy
of the lower level, the logarithm of the oscillator strength times the
statistical weight and the Solar equivalent width.
The Solar abundance adopted from Anders \& Grevesse (1989), except for
iron, is also given for each element.
The Solar iron abundance was adopted from Biemont et~al. (1991) and
Holweger et~al. (1991)
}
\vbox {\halign {
#\hfil&\quad\hfil#&\quad\hfil#&\quad\hfil#&\quad\quad\quad\quad#\hfil&
\quad\hfil#&\quad\hfil#&\quad\hfil#&\quad\quad\quad\quad#\hfil&\quad\hfil#&
\quad\hfil#&\quad\hfil#\cr
\multispan{12}\hrulefill\hfil\cr
\hfil $\lambda$\hfil &$\chi_\ell$\hfil &$\log gf$\hfil &
$W_{\lambda\odot}$\hfil & \hfil $\lambda$\hfil &$\chi_\ell$\hfil &
$\log gf$\hfil &$W_{\lambda\odot}$\hfil & \hfil $\lambda$\hfil &
$\chi_\ell$\hfil &$\log gf$\hfil &$W_{\lambda\odot}$\hfil \cr
\hfil [\AA]\hfil &[eV]\hfil & & [m\AA]\hfil & \hfil [\AA]\hfil &[eV]\hfil & &
 [m\AA]\hfil & \hfil [\AA]\hfil &[eV]\hfil & & [m\AA]\hfil \cr
\multispan{12}\hrulefill\hfil\cr
\multispan3\hfil {Na~{\sc i} ~~$\log \epsilon_\odot = 6.33$} & &
      \multispan3\hfil {Fe~{\sc i} ~~$\log \epsilon_\odot = 7.51$} & &
      ~5909.98&3.21&$-$2.59&38~~\cr
{}~5682.65&2.10&$-$0.77&109~~&                                        ~5461.56&
4.44&$-$1.62&28~~&                                          ~5916.26&2.45&
$-$2.95&57~~\cr
{}~5688.22&2.10&$-$0.60&127~~&                                        ~5462.97&
4.47&$-$0.48&100~~&                                          ~5952.73&3.98&
$-$1.38&65~~\cr
{}~6154.23&2.10&$-$1.63&39~~&                                         ~5463.29&
4.43&$-$0.41&110~~&                                          ~5976.79&3.94&
$-$1.30&72~~\cr
{}~6160.75&2.10&$-$1.33&60~~&                                         ~5501.48&
0.96&$-$3.22&119~~&                                          ~5983.69&4.55&
$-$0.80&71~~\cr
\multispan3\hfil {Mg~{\sc i} ~~$\log \epsilon_\odot = 7.58$} & &      ~5506.79&
0.99&$-$3.14&123~~&                                           ~5984.83&4.73&
$-$0.44&88~~\cr
{}~5711.10&4.35&$-$1.84&112~~&                                        ~5519.58&
3.30&$-$2.70&28~~&                                           ~5987.07&4.79&
$-$0.57&74~~\cr
\multispan3\hfil {Si~{\sc i} ~~$\log \epsilon_\odot = 7.55$} & &      ~5522.45&
4.21&$-$1.51&45~~&                                           ~6024.07&4.55&
$-$0.30&112~~\cr
{}~5665.56&4.92&$-$2.03&43~~&                                         ~5543.20&
3.69&$-$1.57&68~~&                                           ~6027.06&4.07&
$-$1.29&66~~\cr
{}~5684.49&4.95&$-$1.70&64~~&                                         ~5543.94&
4.22&$-$1.16&65~~&                                           ~6042.10&4.65&
$-$0.93&57~~\cr
{}~5772.15&5.08&$-$1.66&59~~&                                         ~5546.51&
4.37&$-$1.21&54~~&                                           ~6078.50&4.79&
$-$0.52&78~~\cr
{}~5793.08&4.93&$-$1.88&52~~&                                         ~5560.22&
4.43&$-$1.16&54~~&                                           ~6079.02&4.65&
$-$1.04&50~~\cr
{}~6145.02&5.61&$-$1.48&40~~&                                         ~5562.72&
4.43&$-$0.99&65~~&                                          ~6089.57&5.02&
$-$0.94&37~~\cr
\multispan3\hfil {Ca~{\sc i} ~~$\log \epsilon_\odot = 6.36$} & &      ~5569.63&
3.42&$-$0.74&159~~&                                          ~6093.65&4.61&
$-$1.40&32~~\cr
{}~5512.99&2.93&$-$0.68&90~~&                                         ~5576.10&
3.43&$-$1.07&120~~&                                           ~6096.67&3.98&
$-$1.85&39~~\cr
{}~5601.29&2.53&$-$0.66&118~~&                                        ~5618.64&
4.21&$-$1.37&53~~&                                           ~6102.18&4.83&
$-$0.37&88~~\cr
{}~6102.73&1.88&$-$1.02&134~~&                                        ~5619.61&
4.39&$-$1.49&37~~&                                           ~6151.62&2.17&
$-$3.36&51~~\cr
{}~6161.29&2.52&$-$1.31&68~~&                                         ~5624.03&
4.39&$-$1.21&53~~&                                           ~6157.73&4.07&
$-$1.32&64~~\cr
{}~6162.18&1.90&$-$0.39&231~~&                                        ~5653.87&
4.39&$-$1.40&42~~&                                           ~6165.36&4.14&
$-$1.57&46~~\cr
{}~6166.44&2.52&$-$1.25&72~~&                                         ~5661.35&
4.28&$-$1.84&25~~&                                           ~6173.34&2.22&
$-$2.95&69~~\cr
{}~6169.04&2.52&$-$0.91&97~~&                                         ~5662.52&
4.18&$-$0.74&97~~&
 \multispan3\hfil {Fe~{\sc ii} ~~$\log \epsilon_\odot = 7.51$} & \cr
{}~6169.56&2.52&$-$0.73&113~~&                                        ~5679.03&
4.65&$-$0.85&62~~&                                           ~6084.10&3.20&
$-$3.84&22~~\cr
\multispan3\hfil {Sc~{\sc ii} ~~$\log \epsilon_\odot = 3.10$} & &     ~5717.84&
4.28&$-$1.08&67~~&                                           ~6149.25&3.89&
$-$2.79&38~~\cr
{}~5657.88&1.51&$-$0.35&71~~&                                         ~5720.90&
4.55&$-$1.72&20~~&
   \multispan3\hfil {Ni~{\sc i} ~~$\log \epsilon_\odot = 6.25$} & \cr
{}~5684.20&1.51&$-$0.92&42~~&                                         ~5731.77&
4.25&$-$1.19&62~~&                                           ~5462.50&3.85&
$-$0.87&47~~\cr
\multispan3\hfil {Ti~{\sc i} ~~$\log \epsilon_\odot = 4.99$} & &      ~5741.86&
4.25&$-$1.64&36~~&                                           ~5614.78&4.15&
$-$0.59&48~~\cr
{}~5953.17&1.89&$-$0.33&36~~&                                         ~5775.09&
4.22&$-$1.15&66~~&                                           ~5625.33&4.09&
$-$0.71&44~~\cr
{}~5965.03&1.88&$-$0.44&31~~&                                         ~5778.46&
2.59&$-$3.42&27~~&                                           ~5682.21&4.10&
$-$0.39&63~~\cr
\multispan3\hfil {Cr~{\sc i} ~~$\log \epsilon_\odot = 5.67$} & &      ~5793.92&
4.22&$-$1.59&40~~&                                           ~5805.23&4.17&
$-$0.59&47~~\cr
{}~5783.07&3.55&$-$0.20&39~~&                                         ~5838.38&
3.94&$-$2.24&22~~&                                           ~6086.29&4.26&
$-$0.55&45~~\cr
{}~5787.93&3.32&$-$0.18&52~~&                                         ~5852.23&
4.55&$-$1.26&42~~&                                           ~6108.13&1.68&
$-$2.56&67~~\cr
\multispan3\hfil {Cr~{\sc ii} ~~$\log \epsilon_\odot = 5.67$} & &     ~5855.09&
4.61&$-$1.56&24~~&                                           ~6111.08&4.09&
$-$0.87&36~~\cr
{}~5502.09&4.17&$-$1.93&21~~&
 ~5856.10&4.29&$-$1.63&35~~&                                          ~6175.37&
4.09& $-$0.60&51~~\cr
 & & & &                                                              ~5859.60&
4.55&$-$0.74&75~~&                                           ~6176.82&4.09&
$-$0.37&65~~\cr
 & & & &                                                              ~5862.37&
4.55&$-$0.55&90~~&
  \multispan3\hfil {Ba~{\sc ii} ~~$\log \epsilon_\odot = 2.13$} & \cr
 & & & &                                                              ~5905.68&
4.65&$-$0.88&60~~&                                             ~5853.69&0.60&
$-$0.94&65~~\cr
\multispan{12}\hrulefill\hfil\cr
}}
\endtab
The adopted line data for all our lines are given in Table~5.
Line identifications, wavelengths and excitation energies were taken from
Moore et~al. (1966).
Van\ der\ Waals broadening parameters were adopted from the same sources as in
EAGLNT.  Since accurate oscillator strengths for moderately weak atomic lines
are not abundantly available, these were derived from the Solar spectrum.
The chemical abundances in the Sun are well known from photospheric and
meteoritic sources (cf. Anders \& Grevesse 1989; Biemont et~al.\ 1991;
Holweger et~al.\ 1991).
Solar equivalent widths were measured from the high quality
Solar Flux Atlas by Kurucz et~al.\ (1984), degraded to the resolution obtained
for our stellar spectra.  We used a similar procedure (as was just described
for
the abundance analysis) to derive the
oscillator strengths of all our spectral lines from the Sun, by varying these
values rather than the abundance, until the Solar model atmosphere yielded line
strengths equal to the measured Solar value for each line.
Such astrophysical oscillator strengths are useful for deriving
chemical abundances relative to the Sun, since the $gf$ values cancel in the
comparison between stellar and Solar abundances.

For our 2 Sc\,{\sc ii} lines Mansour et~al. (1989) give data concerning
atomic hyperfine structure.
The wavelength splitting of the 13 components of the 5657.88 and the 9
components of the 5684.20\,\AA\ lines are, however, less than 25\,m\AA,
and have a negligible effect ($< 0.02$\,dex) on our resulting abundances.

The abundance results and formal statistical errors for the different species
in
the programme stars are given in Table~6.
\begtabfullwid
\tabcap{6} {
Abundance results for the programme stars.
For each star and species ``X'' is given the abundance relative to the Sun
``[X/H]'', the formal standard deviation of the mean ``$\bar\sigma$''
and the number of lines ``n'' used in the determination
}
\vbox {\halign {
#\hfil&
  \hfil#&\hfil#&\hfil#&
\,\hfil#&\hfil#&\hfil#&
\,\hfil#&\hfil#&\hfil#&
\,\hfil#&\hfil#&\hfil#&
\,\hfil#&\hfil#&\hfil#&
\,\hfil#&\hfil#&\hfil#&
\,\hfil#&\hfil#&\hfil#&
\,\hfil#&\hfil#&\hfil#&
\,\hfil#&\hfil#&\hfil# \cr
\multispan{28}\hrulefill\hfil\cr
 No. &
\multispan3\hfil~~~5\hfil&
\multispan3\hfil~~~8\hfil&
\multispan3\hfil~~28\hfil&
\multispan3\hfil~~47\hfil&
\multispan3\hfil~~57\hfil&
\multispan3\hfil~~60\hfil&
\multispan3\hfil~~63\hfil&
\multispan3\hfil~~70\hfil&
\multispan3\hfil~125\hfil \cr
 & \multispan3\,~\hrulefill
 & \multispan3\,~\hrulefill
 & \multispan3\,~\hrulefill
 & \multispan3\,~\hrulefill
 & \multispan3\,~\hrulefill
 & \multispan3\,~\hrulefill
 & \multispan3\,~\hrulefill
 & \multispan3\,~\hrulefill
 & \multispan3\,~\hrulefill \cr
 X &
 {}~[X/H] &$\bar\sigma$~~& n&
 {}~[X/H] &$\bar\sigma$~~& n&
 {}~[X/H] &$\bar\sigma$~~& n&
 {}~[X/H] &$\bar\sigma$~~& n&
 {}~[X/H] &$\bar\sigma$~~& n&
 {}~[X/H] &$\bar\sigma$~~& n&
 {}~[X/H] &$\bar\sigma$~~& n&
 {}~[X/H] &$\bar\sigma$~~& n&
 {}~[X/H] &$\bar\sigma$~~& n \cr
\multispan{28}\hrulefill\hfil\cr
Na\,{\sc i} &        &     &  &  0.20 & .09 & 3&  0.17 & .05 & 3&  0.20  &
&
 1&  0.14  & .10 & 2&  0.05  &     & 1&  0.10  & .08 & 3&  0.19 & .06 & 2&
   0.33 & .03 & 2 \cr
Mg\,{\sc i} &        &     &  &  0.13 &     & 1&$-$0.04 &     & 1&        &
     &  &  0.07  &     & 1&  0.28  &     & 1&  0.06  &     & 1&  0.10 &     &
1&
        &     &   \cr
Si\,{\sc i} &        &     &  &  0.16 & .06 & 4&  0.10 & .07 & 3&  0.15  & .15
&
 3&  0.02  & .08 & 4&$-$0.20 &     & 1&  0.18  & .08 & 4&  0.01 & .03 & 3&
   0.27 &     & 1 \cr
Ca\,{\sc i} &$-$1.16 & .03 & 7&  0.23 & .03 & 8&  0.18 & .03 & 8&  0.31  &
&
 1&  0.31  & .08 & 2&  0.16  & .07 & 3&  0.18  & .04 & 8&  0.25 & .05 & 7&
   0.09 & .10 & 3 \cr
Sc\,{\sc ii}&        &     &  &  0.07 & .16 & 2&  0.13 &     & 1&        &
&
  &        &     &  &        &     &  &  0.02  & .05 & 2&  0.19 &     & 1&
        &     &   \cr
Ti\,{\sc i} &        &     &  &       &     &  &       &     &  &        &
&
  &        &     &  &        &     &  &  0.40  & .05 & 2&       &     &  &
        &     &   \cr
Cr\,{\sc i} &        &     &  &  0.26 &     & 1&       &     &  &        &
&
  &  0.09  &     & 1&        &     &  &  0.11  & .05 & 2&  0.02 & .14 & 2&
   0.38 &     & 1 \cr
Cr\,{\sc ii}&        &     &  &  0.16 &     & 1&  0.01 &     & 1&        &
&
  &        &     &  &        &     &  &  0.18  &     & 1&       &     &  &
        &     &   \cr
Fe\,{\sc i} &$-$1.17 & .06 & 5&  0.23 & .02 &27&  0.10 & .02 &34&  0.25  & .04
&
13&  0.23  & .03 &29&  0.02  & .03 &11&  0.21  & .01 &51&  0.24 & .02 &33&
   0.22 & .03 & 17 \cr
Fe\,{\sc ii}&        &     &  &  0.14 & .01 & 2&  0.07 &     & 1&        &
&
  &  0.15  &     & 1&$-$0.16 &     & 1&  0.15  & .04 & 2&  0.22 & .01 & 2&
   0.17 &     & 1 \cr
Ni\,{\sc i} &$-$0.76 & .10 & 2&  0.13 & .09 & 5&  0.20 & .04 & 4&        &
&
  &        &     &  &        &     &  &  0.10  & .03 & 9&  0.08 & .01 & 2&
   0.10 &     & 1 \cr
Ba\,{\sc ii}&        &     &  &  0.30 &     & 1&  0.11 &     & 1&  0.30  &
&
 1&  0.51  &     & 1&        &     &  &  0.20  &     & 1&  0.26 &     & 1&
   0.31 &     & 1 \cr
\multispan{28}\hrulefill\hfil\cr
}}
\endtab

\titlea {Errors in the abundance results}
\titleb {Errors in the atomic line data}
Errors in the astrophysical $gf$ values may be due to line misidentifications,
undetected line blends, and errors in the solar equivalent width measurements.
These errors are difficult to estimate, but are probably very small for species
with many lines measured.
For species with only one or two lines measured they may be as large as
0.1\,dex.

Since the programme stars are hotter than the Sun, and many lines weaker, also
errors in the van\ der\ Waals damping parameters may affect the resulting
abundances relative to the Sun.
As a test we repeated the full analysis (both the $gf$ values determination and
the abundance analysis) of No.~70 without any enhancement factors to the
classical damping constants (cf EAGLNT).
This resulted in decreased stellar abundances by between 0.01\,dex for
Fe\,{\sc ii} and 0.11\,dex for Mg\,{\sc i} and Ca\,{\sc i}.
The classical damping constants are almost invariably found to be too small,
and we conclude that the likely uncertainties in the van\ der\ Waals damping
cause abundance errors smaller than 0.05\,dex for all species.

\titleb {Errors in the fundamental atmospheric parameters}
\titlec {Uncertainties in the $uvby$ photometry}
Due to the relative faintness of the stars and the different sources of
photometry we expect comparatively large random uncertainties in the model
parameters:
The uncertainties in $b-y$, $m_1$ and $c_1$ are estimated to be 0.015, 0.020
and 0.020 magnitudes, respectively, based on comparisons between the
$uvby\beta$ photometric data
found in WGLPB, Epstein (1968), Eggen (1972) and Graham \& Slettebak (1973).
These can be translated into errors in the model parameters: 120\,K in
$T_{\rm eff}$, 0.2\,dex in $\log~g$, 0.25\,dex in [Me/H] and
0.4\,km\,s$^{-1}$ in $\xi_{\rm t}$.
For Nos.~47 and 125 we expect the errors to be about 50\% larger due to the
approximate methods used to derive their atmospheric parameters.
\titlec {Microturbulence parameters and effective temperatures}
As mentioned in section 2.3 many of the lines used in the abundance analysis
fall on the microturbulence sensitive part of the curve of growth.
In a first analysis the microturbulence parameters of the programme stars were
assumed to follow the relation derived by EAGLNT for Pop~I field dwarfs,
but the resulting abundances showed large internal inconsistencies:
A plot of abundance vs. equivalent width for a well represented species (here:
Fe\,{\sc i})
for a star yields a test of the adopted microturbulence parameter.
An erroneous value of $\xi_t$ would result in a slope in the diagram if all
other parameters could be assumed correct.
An example of such a diagram for two values of the microturbulence parameter
can be seen in Fig.~2a and b.
\begfig 11.2 cm
\figure {2}{
{\bf a} and {\bf b}: abundances derived from Fe\,{\sc i} lines vs. equivalent
width, $W_\lambda$, for No.~70 with two different assumptions of the
microturbulence parameter.  Linear least squares fits and their slopes are also
shown.  In panel {\bf c} the effect of a change in the effective temperature
relative to {\bf a} is shown.  To avoid systematic errors the equivalent widths
are not the measured ones, but the so called theoretical equivalent widths,
cf.\ Magain (1984).  {\bf d, e} and {\bf f} shows abundance [Fe/H] vs. line
excitation energy for the same choices of microturbulence and effective
temperature.  The symbols are selected to facilitate identification of lines
between the left and right panels:
The filled symbols show lines with $\chi_\ell < 1.0$\,eV, open symbols are
lines
with excitation energies between 2.1 and 2.6\,eV and the starred symbols show
lines with $\chi_\ell > 3.4$\,eV (cf {\bf d}.
As seen in {\bf a}, symbols with 3 vertices show lines with equivalent widths
$<~45$\,m\AA, symbols with 10 vertices show lines with equivalent widths
$> 100$\,m\AA\ and symbols with 4 vertices show lines with intermediate
equivalent widths
}
\endfig
Fig.~2a, b and c shows that the slopes found are not very significant, which
is true for the whole sample of stars; the inclusion or exclusion of one or two
lines may change the slopes.
Nevertheless, the mean of the slopes for several stars is judged to be more
significant and we conclude that the microturbulence parameters for the metal
rich stars should in the mean be increased by 0.5\,km\,s$^{-1}$, a change which
was adopted in the final abundance analysis.
The modified values are shown in Table~4.

The reason why our cluster stars appear to require higher microturbulence
parameters than a sample of F field dwarfs which have already begun to evolve
away from the ZAMS may be sought in the age difference.
It is possible that young, rather quickly rotating stars, having (probably)
high
chromospheric activity, are more turbulent even in their photospheres.

Panel c of Fig.~2 shows that the diagram is also affected by a change of the
effective temperature of the star, demonstrating that the available sample of
Fe\,{\sc i} lines is not ideal for independent testing of our determinations of
microturbulence parameters and effective temperatures since there is a
correlation between excitation energy and line strength in our sample of lines
in these stars.

We conclude from this discussion that a reasonable random uncertainty in the
microturbulence parameters is 0.5\,km\,s$^{-1}$.

By investigating whether the abundances derived from several Fe\,{\sc i} lines
of different excitation energies are independent of line excitation energy, the
effective temperatures can be checked.
An example of this can be seen in Fig.~2d and f, where abundance
is plotted vs. excitation energy, $\chi_\ell$, for two choices of $T_{\rm
eff}$.
The figure suggests that the effective temperature for No.~70 should be
lowered by about 100\,K.
Comparison with panel e also shows the coupling of the slope with the
microturbulence parameter.
The changes thus suggested to the photometrically derived temperatures are
shown
in Table~7.  These changes are surprisingly large in view of the above
estimated
photometric uncertainties, and are not seen in our comparison stars discussed
in
5.2.
\begtabfull
\tabcap{7} {
Changes to the photometric model parameters suggested by the spectroscopic
consistency checks.  We argue in the text that the large temperature
corrections
in the middle column are spurious.  (For Nos.~5 and 47 no Fe\,{\sc ii} lines
were measured and for No.~125, no low or medium excitation Fe\,{\sc i} lines
were measured)
}
\vbox {\halign {
#\hfil&\hfil#&\hfil# \cr
\multispan{3}\hrulefill\hfil\cr
No &~~~~~$\Delta T_{\rm eff}$ [K]~&~~~~~$\Delta \log g$~\cr
\multispan{3}\hrulefill\hfil\cr
5   & $-$230 ~~ &         ~~\cr
8   & $-$240 ~~ &   +0.30 ~~\cr
28  &  $-$40 ~~ &   +0.10 ~~\cr
47  & $-$580 ~~ &         ~~\cr
57  & $-$570 ~~ &   +0.25 ~~\cr
60  &  $-$80 ~~ &   +0.55 ~~\cr
63  & $-$260 ~~ &   +0.20 ~~\cr
70  &  $-$90 ~~ &   +0.07 ~~\cr
125 &        ~~ &   +0.07 ~~\cr
\multispan{3}\hrulefill\hfil\cr
}}
\endtab

We see at least three reasons why we should not trust these large temperature
errors: \item{1.} $T_{\rm eff}$($\beta$): Effective temperatures may also be
derived from the reddening insensitive H$_\beta$ photometric index.
Such a calibration was made by Saxner \& Hammarb\"ack (1985), and application
to the $\beta$ photometry of WGLPB results in temperatures $125 \pm 115$\,K
cooler than those given in Table~4.
This is in good agreement with the results of EAGLNT who show that the
temperature scale adopted here is about 150\,K hotter than that of
Saxner \& Hammarb\"ack in this range of $T_{\rm eff}$ and metallicities.
A possible systematic uncertainty in this $T_{\rm eff} - \beta$ calibration,
however, is that the $\beta$ indices of our young cluster stars may be affected
by chromospheric activity.  \item{2.} The medium excitation energy lines:
The lines seen in Fig.~2 have been divided into three groups according to
excitation energy: low: two lines at 0.96--0.99\,eV,
medium: four lines  at 2.17--2.59\,eV
and high: many lines at 3.2--5.1\,eV.
(The three groups are shown with filled, open and starred symbols,
respectively,
in Fig.~2.)
For most stars the large slopes of the linear fits, suggesting the large
temperature corrections in Table~7, are governed by the very high abundances
derived for the two low excitation lines.
The medium excitation lines, on the other hand, yield abundances typical for
the
high excitation lines, and not intermediate values as one might expect if the
adopted effective temperatures were indeed too high.
\item{3.} The microturbulence sensitivity of the low excitation lines:
The high abundances derived from these two lines may in fact be caused by still
underestimated microturbulence parameters.
The effect of a reasonable change to this fitting parameter can be seen in
panel b of Fig.~2.
In fact, an increase of $\xi_{\rm t}$ by 0.4\,km\,s$^{-1}$ changes the slope in
the diagram as much as an effective temperature change of $-$400\,K.

\noindent
Thus, despite the temperature errors suggested from plots like Fig.~2 we
suggest
that the effective temperatures are correct within the photometric
uncertainties
and that the values in the first column of Table~7 are misleading.

Table~8 shows the sensitivities of the abundance results to changes in the
effective temperatures, surface gravities and microturbulence parameters.
\begtabfull
\tabcap{8} {
Sensitivities of the abundance results to changes in the model parameters.
Stars No.\,8 and No.\,70 were used as test cases for deriving the table
}
\vbox {\halign {
#\hfil&\hfil#&\hfil#&\hfil# \cr
\multispan{4}\hrulefill\hfil\cr
Species &$\Delta T_{\rm eff}=$+200~K~&$\Delta \log g=$+0.2~dex~&
$\Delta \xi_{\rm t}=$+0.4~km~s$^{-1}$~ \cr
\multispan{4}\hrulefill\hfil\cr
{}~Na~{\sc i} &   +0.10~~~~~~ & $-$0.03~~~~~~~~~ & $-$0.03~~~~~~~~~~ \cr
{}~Mg~{\sc i} &   +0.12~~~~~~ & $-$0.02~~~~~~~~~ & $-$0.05~~~~~~~~~~ \cr
{}~Si~{\sc i} &   +0.08~~~~~~ & $-$0.01~~~~~~~~~ & $-$0.02~~~~~~~~~~ \cr
{}~Ca~{\sc i} &   +0.14~~~~~~ & $-$0.04~~~~~~~~~ & $-$0.07~~~~~~~~~~ \cr
{}~Sc~{\sc ii}&   +0.02~~~~~~ &   +0.05~~~~~~~~~ & $-$0.16~~~~~~~~~~ \cr
{}~Cr~{\sc i} &   +0.08~~~~~~ & $-$0.01~~~~~~~~~ & $-$0.03~~~~~~~~~~ \cr
{}~Cr~{\sc ii}&               &   +0.05~~~~~~~~~ &                   \cr
{}~Fe~{\sc i} &   +0.12~~~~~~ & $-$0.01~~~~~~~~~ & $-$0.07~~~~~~~~~~ \cr
{}~Fe~{\sc ii}& $-$0.04~~~~~~ &   +0.05~~~~~~~~~ & $-$0.09~~~~~~~~~~ \cr
{}~Ni~{\sc i} &   +0.14~~~~~~ &    0.00~~~~~~~~~ & $-$0.07~~~~~~~~~~ \cr
{}~Ba~{\sc ii}&   +0.06~~~~~~ &   +0.04~~~~~~~~~ & $-$0.22~~~~~~~~~~ \cr
\multispan{4}\hrulefill\hfil\cr
}}
\endtab
\titlec {Ionization balances -- $\log~g$}
The ratio between the abundances derived from neutral and ionized species of
the
same element is mainly sensitive to the assumptions of $T_{\rm eff}$ and
$\log\,g$ via their effects on the ionization balances.
Table~6 shows that the resulting abundance ratios Fe\,{\sc i}/Fe\,{\sc ii} and
Cr\,{\sc i}/Cr\,{\sc ii} are larger than one in almost all cases.
Table~8 shows typical abundance effects of changes to $\log\,g$, and
Table~7 shows the changes to the surface gravities that would satisfy the
ionization balances (assuming the other model parameters to be correct):
1typically +0.2\,dex.  We conclude that the abundance effects of these errors
are less than 0.05\,dex.  Lowering of the effective temperatures by 70\,K would
have similar effects on the ionization balances.
\titlec {Overall metallicities}
The model atmospheres were first calculated with metallicities derived from the
Str\"omgren $m_1$ index (c.f. Sect.~3.2.1 and Table~4).
The final abundances were derived with new models with metallicities derived in
a preliminary analysis, but the results were found to be quite insensitive to
reasonable errors in this parameter.
\titlec {Unresolved binaries}
It is likely that our sample of stars contains unresolved binaries.
An analysis assuming that a binary is a single star may then give erroneous
results.
We have made an investigation of one possible such case, assuming an F5V
primary with a G5V companion.
According to Allen (1973) such stars at the ZAMS have the following approximate
characteristics:
F5V: $M_V=3.9$, $B-V=0.42$, $T_{\rm eff}=6580$\,K, $M/M_\odot=1.3$,
and
G5V: $M_V=5.2$, $B-V=0.70$, $T_{\rm eff}=5520$\,K, $M/M_\odot=0.9$.
The $b-y$ colours would be approximately 0.285 and 0.43, respectively.
Combining these would yield a binary with $M_V=3.6$, $B-V=0.48$,
$b-y=0.32$ and $T_{\rm eff}(b-y)=6400$\,K, which is 180\,K less than that of
the primary.
The flux ratio $F_{\rm A}$/$F_{\rm B}$ in the observed region is about 2.
We have calculated equivalent widths for the lines in our investigation for two
models of these descriptions, assuming $\log\,g=4.5$ and [Fe/H]=+0.2.
The equivalent widths of neutral lines turn out to be 50--80\% larger in the
secondary than in the primary component, and for the 2 Fe\,{\sc ii} lines the
widths are 30\% smaller in the secondary.
The combination of spectra of two such components suggest that the equivalent
widths of neutral lines in our investigation are typically 20\% larger in the
binary than in the primary component alone, while the Fe\,{\sc ii} lines are
weakened by about 10\%.
A look at Table~8 shows that the lower $T_{\rm eff}$ of the binary relative to
the primary lowers the abundances of the neutral lines by 0.06--0.11\,dex,
while
the strengthening of the line widths should increase the abundances by a
similar
amount.
For the Fe\,{\sc ii} lines, both the $T_{\rm eff}$ and line-strength effects go
in the opposite directions, as compared to Fe\,{\sc i}, but they also largely
cancel in the resulting abundances.
For the Ba\,{\sc ii} and Sc\,{\sc ii} lines the equivalent widths change very
little and the abundances are thus somewhat too low if the binary is treated
like a single star.

We conclude that the effects of binarity on our abundances are likely to be
small.
\titleb {Comparison with other results}
In order to check the data reductions, the method of analysis and to get an
estimate of the uncertainties in our abundance results, four stars analysed by
EAGLNT were also observed with CASPEC and analysed by our methods.
The abundance results of the two analyses are compared in Fig.~3.
\begfig 8.0 cm
\figure {3}{
Abundances derived for our four comparison stars; filled symbols show the
results derived here, and open symbols show abundances derived by EAGLNT
}
\endfig
It should be noted that the line half-widths of these ``standard stars'' are,
with one exception, smaller than those of our programme stars due to the
relative youth of the latter with considerable rotational line broadening
(cf.\ Tables~1 and 2).

There seem to be some systematic differences between the results of the two
investigations.
First we see that the EAGLNT abundances relative to the solar abundances
(open symbols in Fig.~3) show rather smooth variations with atomic number,
while our results seem to scatter more.

For HR\,33, HR\,7232 and HR\,8181 our abundance values are typically
0.05--0.10\,dex (10--25\%) higher than those of EAGLNT.
These three stars are considerably cooler than our programme stars.
For HR\,7126, which is more like our programme stars as regards the model
parameters and line-widths, we derive instead typically 0.05\,dex (10\%) lower
abundances than EAGLNT.

Since the methods of analysis and the model atmospheres are identical the
differences have to be in the equivalent width measurements or due to
differences between the line data sets.
The line data sets of our investigation have 25 lines in common with that of
EAGLNT in the wavelength region 5800--6170 \AA.
For the lines in common between the two investigations our equivalent widths
for HR\,33 and HR\,232 are typically 6\,m\AA\ higher than those measured from
high resolution spectra by EAGLNT (cf. Edvardsson et~al. 1993b).
For HR\,8181 our widths are typically 2.5\,m\AA\ higher and
for HR\,7126 our values are about 6\,m\AA\ lower.
Since many of the lines fall on the ``flat'' part of the curve of growth we do
not expect a direct proportionality between equivalent width and derived
abundance, but the agreements between the signs of the differences in
abundances
and equivalent widths clearly suggest that there are systematic differences in
the continuum levels of the two investigations.
These are probably explained by the lower resolution of our CASPEC spectra
which will also increase the likelihood that line blends will go undetected,
resulting in systematic overestimates of the equivalent widths.

It is interesting that HR\,7126, which is quite similar to our metal rich stars
both as regards the effective temperature and the FWHM of the spectral lines,
shows lower abundances in our investigation as compared to EAGLNT, while the
three cooler, narrow-lined stars show the opposite behaviour.
We therefore suspect that our results for the metal rich stars may be
underestimated by typically 0.05\,dex (10\%) compared to what would have been
found in a study of high resolution spectra.

For the metal poor, cooler, narrow-lined field star No.~5 we may, on the other
hand, have overestimated the abundances by 0.1\,dex (25\%) compared to what
might have been found in a study with higher spectral resolution.
\titleb {Summary of the error discussion}
\titlec {Random errors}
The previous discussion of errors in the model parameters lead to the following
error estimates: standard deviations; $T_{\rm eff}$ $\pm$120\,K, $\log\,g$
$\pm$0.2\,dex and $\xi_t$ $\pm$0.5\,km\,s$^{-1}$.
With the help of Table~8 these uncertainties, when assumed uncorrelated, are
found to generate uncertainties of 0.10\,dex for a typical element.

Addition of these errors to the statistical errors in Table~6, give at hand
that
we may expect a random standard deviation error of about $\pm$0.10--0.12\,dex
for an element with many lines analysed, and somewhat larger for less well
represented elements.
\titlec {Systematic errors}
As discussed in EAGLNT, there may be systematic errors in the photometric
effective temperature calibration of $-$50 to +100\,K, corresponding to
abundance
uncertainties of typically $-$0.03 to +0.06\,dex for lines of neutral species.
Furthermore, we note from the discussion in Sect.~4.1.2 that our
microturbulence
parameters for the metal rich stars may still be systematically too low by
$\approx$0.5\,km\,s$^{-1}$, which would cause us to overestimate the abundances
by about 0.07\,dex.
Our surface gravities may be too low by typically 0.2\,dex which results in
only
small errors in the abundance analysis.
Finally the uncertainties in the equivalent width measurements, probably caused
by the uncertainty in the continuum level in our spectra, may cause us to
systematically underestimate the abundances for the cluster members by
0.05\,dex.

Adding these maximum errors we get a maximum systematic error in the cluster
abundances of $\pm 0.11$\,dex for neutral species and, with a similar analysis,
+0.10 to $-0.20$\,dex for ionized species.

\titlea {Discussion}
\titleb {Cluster membership}
We have three entirely independent criteria to determine cluster membership for
our programme stars: radial velocities, distance moduli and chemical
abundances.
Star No.~5 is definitely not a member with its Population II chemical abundance
and high radial velocity, and with a distance modulus being smaller than those
of the others; it is not considered in the following discussion.
For four stars, Nos.~47, 63, 70 and 125; all the three criteria indicate
membership.
This is the group that we use to derive the cluster mean parameters:
$\overline{RV}=3.9 \pm 0.7$, $\langle V_{\rm 0}-M_V\rangle=6.92 \pm 0.01$
and chemical abundances according to Table~9.
\begtabfull
\tabcap{9} {
Mean chemical abundances relative to Solar, internal standard deviation of the
mean, $\bar\sigma$, and total number of lines, n, used for four assumed members
of the $\zeta$\,Sculptoris cluster.
The last column, p.e., gives an estimated total probable error and is based on
internal errors, systematic errors and the number of lines in the mean value
}
\vbox {\halign {
#\hfil&\hfil#&\quad\hfil#&\quad\hfil#&\quad\hfil# \cr
\multispan{5}\hrulefill\hfil\cr
Species X & ~~~~~[X/H] & ~~~~$\bar\sigma$ & ~~~~n & ~~~~p.e. \cr
\multispan{5}\hrulefill\hfil\cr
{}~~Na~{\sc i} &$+$0.19        & .04 &   8 & .15\cr
{}~~Mg~{\sc i} &$+$0.08\rlap : & .03 &   2 & .20\cr
{}~~Si~{\sc i} &$+$0.13        & .05 &  11 & .10\cr
{}~~Ca~{\sc i} &$+$0.19        & .03 &  19 & .10\cr
{}~~Sc~{\sc ii}&$+$0.06\rlap : & .08 &   3 & .20\cr
{}~~Ti~{\sc i} &$+$0.41\rlap : & .05 &   2 & .20\cr
{}~~Cr~{\sc i} &$+$0.13\rlap : & .08 &   5 & .15\cr
{}~~Cr~{\sc ii}&$+$0.18\rlap : &     &   1 & .20\cr
{}~~Fe~{\sc i} &$+$0.23        & .01 & 114 & .10\cr
{}~~Fe~{\sc ii}&$+$0.18\rlap : & .02 &   5 & .15\cr
{}~~Ni~{\sc i} &$+$0.10        & .02 &  12 & .10\cr
{}~~Ba~{\sc ii}&$+$0.27\rlap : & .03 &   4 & .15\cr
\multispan{5}\hrulefill\hfil\cr
}}
\endtab
The cluster distance from the distance modulus is 242\,pc with a
standard deviation of only 2\,pc, which is less than the apparent cluster
radius of about $1\degr$ on the sky; 4\,pc at that distance.
Of the remaining four stars for which we have chemical abundances, No.~8 has
0.32\,mag smaller distance modulus, corresponding to about 35\,pc smaller a
distance than the others.
This may be caused by multiplicity, but the radial velocity of No.~8 appears to
be constant, possibly the orbital plane is near the plane of the sky as
discussed in Sect.~3.1.
No.~28 has a velocity and distance modulus in agreement with the quartet
above, but its (Fe\,{\sc i}) abundance is down by 0.1\,dex. This difference is
probably significant as it is seen also in the data of WGLPB and therefore it
is not included in the determination of the mean cluster parameters.
No.~57 has quite a deviating but variable radial velocity
($\overline{RV}= -21 \pm 5$\,km\,s$^{-1}$ from only two
observations), which may be due to duplicity.
For this star more radial velocity determinations are desirable.
For No.~60 we derive about 0.2\,dex lower iron (Fe\,{\sc i}) abundance and
0.05\,dex lower calcium abundance than the quartet.
For this star we could only measure a few lines due to the rotational line
broadening and the scattered results for one line each of Na, Mg, Si and
Fe\,{\sc ii} are very uncertain.
Our discussion of the effects of line broadening for the continuum
level for the case of HR\,7126 in Sect.~4.3 suggests that the abundances may
actually be as high as for the quartet.
\titleb {Cluster chemical abundances}
Our finding that the cluster is metal rich contrasts with the photometrically
derived metallicities of WGLPB:
from the $UBV$ photometry a cluster metallicity ${\rm [Me/H]} = -0.21 \pm 0.05$
may be found for ``our'' members Nos.~47, 63, 70 and 125.
The WGLPB results from the $uvby\beta$ photometry for these stars (except
No.~47
and 125 for which the Str\"omgren photometry was rejected or not given) is
${\rm [Me/H]} = +0.24 \pm 0.11$.
WGLPB noted an ultraviolet excess of about 0.1\,mag in $U$ in the $U-B$ vs
$B-V$
diagram and, perhaps, 0.05\,mag in Str\"omgren $u$ in the $u-b$ vs $b-y$
diagram, with respect to standard relations for dwarf stars.
The UV excess(es) is supposedly due to stronger stellar activity in these young
stars as compared to the stars defining the standard relations.
This ``extra'' UV excess may well explain the low metallicity found from the
$UBV$ photometry, since the $U$-passband excess is interpreted as less line
blanketing, implying too low a metallicity.
The ``Str\"omgren'' metallicity is in principle derived from the line
blanketing
in the $v$ passband, with its short wavelength edge near 3900\,\AA.
It should therefore be less affected by the UV excess, but the standard
relation
is normalized by the H$\beta$ index (near 4861\,\AA), which could well be
affected by stellar activity.
We find, however, no inconsistency when we compare our effective temperatures
directly derived from $b-y$ with $T_{\rm eff}(\beta)$ in Sect.~4.1.2.

The microturbulence parameters are found to be about 0.5\,km\,s$^{-1}$
higher than in older stars of similar description.
We speculate that this effect may have the same cause as the UV excesses; young
stars are more active, probably because of more turbulent atmospheres and
stronger magnetic fields.

\begfig 4.1 cm
\figure {4}{
Comparison of mean abundances derived for the $\zeta$\,Scl cluster
(filled symbols) with abundances of EAGLNT disk dwarfs having
[Fe/H]$\;\ge +0.15$ (open symbols).
The lines from the open symbols indicate the abundance ranges for the EAGLNT
stars (i.e. not the standard deviations).
The 10 EAGLNT stars have almost the same mean iron and calcium abundance as
$\zeta$\,Scl
}
\endfig
To enable a comparison with other metal rich objects, the chemical abundances
of our four probable $\zeta$\,Scl members are compared to those of the 10 most
metal rich stars of the EAGLNT sample in Fig.~4.
These 10 EAGLNT stars have $+0.15 \le $[Fe/H]$_{\rm Fe~I} \le +0.26$,
with a mean value of [Fe/H] close to our value for $\zeta$\,Scl.
They were formed in different regions and at different times in the galactic
disk, and they are all significantly older than our cluster;
Their ages range from 1.6 to 5.3\,Gyr, and their maximum distances from the
galactic disk, $Z_{\rm max}$, range from 10 to 410\,pc, with $Z_{\rm max}$
larger than 70\,pc only for 3 stars older than 3.5\,Gyr.
(In fact: none of the 20 EAGLNT stars with ages less than 2.5\,Gyr have
$Z_{\rm max} \ge 200$\,pc.)

Starting the comparison with the lighter elements, Na and Ca make no surprises.
The differences seen for Mg and Ti are probably not significant considering our
uncertainties.
Silicon, on the other hand, may well be underabundant.

The most interesting observation is that the Ni abundance is significantly
lower, relative to iron, in $\zeta$\,Scl than in any one of the 10 EAGLNT
stars.
The Ni/Fe abundance ratio is only $82 \pm 7$\% of the solar ratio while it is
larger than the solar ratio for all the 10 metal rich stars of the EAGLNT
sample.
This abundance ratio is quite insensitive to random and systematic errors
in the analyses since the atomic properties and details of the line formation
and the line data sets for Fe\,{\sc i} and Ni\,{\sc i} are very similar,
and the number of lines analysed is comfortable for both species.
Such low Ni/Fe ratios may possibly be produced in some Type~I and II
supernovae,
although iron peak element production in supernovae still is subject to large
uncertainties due to the complexity of the hydrodynamics calculations
(Thielemann 1993).
Khokhlov et~al. (1993) list calculations where some detonating white dwarfs,
producing supernovae of Type~I, make Ni/Fe ratios down to 75\% of solar.

The Ba/Fe abundance ratio, finally, may be somewhat higher than typical for
the comparison stars.
This agrees with the EAGLNT result that the interstellar medium Ba abundance
increases more rapidly with time than the Fe abundance in the galactic disk.

We conclude that there is at least one significant difference in the abundance
pattern deduced for the $\zeta$\,Scl cluster relative to other metal rich stars
in the solar neighbourhood.
This may possibly be related to the youth of the $\zeta$\,Scl cluster.
The difference may also be connected with the unusually high galactic latitude
of the cluster and may tell about an unusual formation history.
\titleb {Cluster origin}
The cluster heliocentric radial velocity was found above to be
$+3.9$\,km\,s$^{-1}$ and the galactic latitude is $-$79\degr.
Therefore the vertical, $W$, component of the cluster velocity may be taken to
be minus (since we are looking south) the heliocentric radial velocity plus the
solar peculiar $W$ velocity which is $+6$\,km\,s$^{-1}$
(M. Grenon, private communication, cf. EAGLNT).
The resulting $\zeta$\,Scl $W$ velocity is only about $+2$\,km\,s$^{-1}$.
This is close to the independent estimate of Eggen (1972) of
$W = -3$\,km\,s$^{-1}$.
A simple dynamic model with the disk as a homogeneous density slab suggests
that
the cluster is oscillating in the $Z$ direction and that it is now
very close to its maximum distance from the galactic plane,
$Z_{\rm max}$, which is therefore adopted to be the cluster distance of
240\,pc.

The typical solar neighbourhood vertical oscillatory period for galactic disk
objects is 62\,Myr (Binney \& Tremaine 1987).
This suggests that the cluster passed through the disk from north to south
about
15\,Myr ago and from south to north about 45\,Myr ago.
Since this is close to the turn-off age 50\,Myr of the cluster
(Perry et~al.\ 1978; de~Epstein \& Epstein 1985)
we suggest that it was actually formed close to the galactic disk 45\,Myr ago
by
processes triggered by a high velocity cloud (HVC) ramming into the galactic
disk interstellar medium, perhaps part of a giant molecular cloud.
This hypothesis may explain the large distance of the cluster from the
galactic plane.
The effects of such an impact have recently been discussed by
Comer\'on \& Torra (1993, cf also references therein).
They support the hypothesis that this process could account for Gould's Belt,
and suggest that it was formed about 40\,Myr ago by a HVC hitting
the interstellar medium in the disk from the southern galactic hemisphere!
Furthermore they suggest that most of the HVC gas would form a star cluster
with a large dynamic scale height, while the shocked interstellar medium
(ISM) would form a ring-like structure with many of the characteristics of
Gould's Belt.  We therefore propose the possibility of a common origin for
the $\zeta$\,Sculptoris open cluster and Gould's Belt.

The somewhat peculiar chemical composition of the cluster may thus be
inherited from gas with an origin very far from the solar neighbourhood.
To deduce the origin of this hypothetical cloud and the possible connection
with
Gould's Belt we need to study the cluster motions further:
A comparison with the kinematic and orbital data of the EAGLNT stars shows that
for objects with Z$_{\rm max} = 240$\,pc the $W$ velocities are about
22\,km\,s$^{-1}$ when passing through the plane of the disk.
Depending on the fraction, $f$, of the cluster mass originating from
the cloud the $W$ velocity of the cloud should have been at least
a factor $1/f$ higher than 22\,km\,s$^{-1}$ when it hit the disk.
Here we have only been able to discuss the vertical velocity component of the
system, which according to Binney \& Tremaine (1987) is quite independent of
the velocity components parallel to the plane of the disk.
To get the full space motion and orbit of the cluster, and, to some
approximation, the direction of motion of the infalling cloud, we need also
proper motion data.
We expect to possess such data in 1995, when the
HIPPARCOS satellite data has been finally reduced.
That data should, of course, also give an independent distance estimate for the
cluster.

An independent check that the cluster is young may be obtained if it is shown
that the lithium abundances for our cluster members are close to the
current ISM abundance:
$\log N_{\rm Li}/N_{\rm H} \approx -9$. The analysed stars are in the effective
temperature range where lithium is destroyed by a still unknown process
after about 100\,Myr (Boesgaard et~al.\ 1988).
Then again, with a possibly unusual origin of the cluster and the
chemical peculiarities suggested above, the Li abundance may not honour
our expectations.  This should be investigated.

The total mass of the cluster was estimated by
de\,Epstein \& Epstein (1985) to about 150\,M$_\odot$.
Assuming a star formation efficiency of 5\% (eg.\ Shu et~al.\ 1987) the
original
mass of the cloud which became the cluster was of the order of 3000\,M$_\odot$,
corresponding to 0.1--10\% of the very uncertain masses of observed HVC:s
(Comer\'on \& Torra 1993).

\titlea {Conclusions}
\item{1.} There is no evidence in the spectral data for a low metallicity in
the young cluster $\zeta$\,Scl.
The cluster seems to be metal rich rather than metal poor,
the average [Fe/H] of 4 probable cluster members being $+0.23$.

\item{2.} Care must be exercised before interpreting $UBV$ photometric
data as metallicity effects if the stars are relatively young.

\item{3.} The chemical abundance pattern in the cluster deviates from
that of a sample of field disk dwarfs with the same mean Fe and Ca abundances:
The nickel abundance is very significantly low, silicon may also be
underabundant.
These differences may be related to the youth of the cluster (about 45\,Myr)
as compared to 1.6 -- 5.3\,Gyr for the comparison field dwarfs.
Additionally, it may be connected to the history of formation of the cluster
which suggests that it may be composed of a mixture of solar neighbourhood gas
and gas from a high-velocity cloud of unknown origin.
The determination of lithium abundances for our programme stars would be
interesting, since present-day ISM abundances could be expected in member
stars with effective temperatures in the lithium-gap range only if the cluster
is as young as supposed.

\item{4.} The microturbulence parameters derived from a set of Fe\,{\sc i}
lines
are systematically higher than those derived for similar but older field
dwarfs.
This may be connected with the youth of the stars with higher turbulent
velocities.

\item{5.} Of the 9 stars with a complete data set (radial velocity, abundance
and distance modulus) four (47, 63, 70 and 125) were selected as representative
for the cluster. The remaining five were discarded either because of deviating
radial velocity (5, 57) abundance (5, 28, 60) or distance (5, 8). This does
not,
however, exclude all of them as possible members, see discussion in Sect.~5.1.
For the stars 38,
54 and 121 we were unable to determine abundances becuse of rotational line
broadening, however both radial velocity data and distance modulus suggests
that
at least 38 remains a strong candidate for cluster membership.

\item{6.} Radial velocity data from 4 assumed cluster members gives a mean
heliocentric radial velocity for the cluster of +3.9\,km s$^{-1}$.
Addition of the solar peculiar velocity shows that the cluster is very close to
its maximum distance from the galactic disk ($\approx 240$\,pc).

\item{7.} We suggest that the cluster was formed 45\,Myr ago by the interaction
of a high velocity cloud (falling from the southern galactic hemisphere)
with gas in the galactic disk.
We also suggest that the age of the cluster, its position in the solar
neighbourhood and the hemisphere-of-origin of the hypothetical infalling cloud,
may indicate an origin in common with Gould's Belt.
The space motion of the cluster should be investigated so that these hypotheses
may be further investigated.

\item{8.} The star CoD\,$-$29\degr18906 (No.\ 5 in WGLPB), is shown to be a
field halo star with a metal abundance [Fe/H]$\,= -1.2$.

\acknow{
We wish to thank H.\,Lindgren for obtaining CORAVEL observations of our
programme stars, and also J.\,Andersen, M.\,Mayor and J.C.\,Mermilliod for
communicating CORAVEL data.
We also thank B.\,Gustafsson for valuable discussions and comments to the
manuscript and F.-K.\,Thielemann for valuable comments of the nuclear yields
of supernovae.
BE acknowledges research grants from the
Swedish Natural Science Research Council, NFR.}

\begref{References}
\ref Allen, C.W. 1973, Astrophysical Quantities 3rd ed., Athlone Press,
Univ. of London
\ref Anders, E., Grevesse, N. 1989, Geochimica et Cosmochimica Acta, 53, 197
\ref Biemont, E., Baudoux, M., Kurucz, R.L., Ansbacher, W., Pinnington, E.H.,
1991, A\&A 249, 539
\ref Binney, J., Tremaine, S. 1987, Galactic Dynamics,
Princeton University Press, 17
\ref Boesgaard, A.M., Budge, K.G., Ramsay, M.E. 1988, ApJ 327, 389
\ref Comer\'on, F., Torra, J. 1993, A\&A 281, 35
\ref Edvardsson, B., Andersen, J., Gustafsson, B., Lambert, D.L., Nissen, P.E.,
Tomkin, J. 1993a, A\&A 275, 101 (EAGLNT)
\ref Edvardsson, B., Andersen, J., Gustafsson, B., Lambert, D.L., Nissen, P.E.,
Tomkin, J. 1993b, A\&AS, 102, 603
\ref Eggen, O.J., 1972, ApJ 173, 63
\ref Epstein, I., 1968, AJ 73, 556
\ref de\,Epstein, A.E.A., Epstein, I., 1985, AJ 90, 1211
\ref Graham ,J.A., Slettebak, A. 1973, AJ 78, 295
\ref Holweger, H., Bard, A., Kock, A., Kock, M., 1991, A\&A 249, 545
\ref Khokhlov, A., M\"uller, E., H\"offlich, P. 1993, A\&A 270, 223
\ref Kurucz, R.L. 1989, Magnetic tapes with atomic and ionic line data,
private communication
\ref Kurucz, R.L., Furenlid, I., Brault, J., Testerman, L. 1984, Solar Flux
Atlas from 296 to 1300\,nm, National Solar Observatory, Sunspot, New Mexico
\ref Lindgren, H. 1993, private communication
\ref Mansour, N.B., Dinneen, T., Young, L., Cheng, K.T. 1989,
Physical Review A 39, 5762
\ref Magain, P. 1984, A\&A 134, 189
\ref Mermilliod, J.C., Mayor, M. 1993, private communication
\ref Moore, C.E., Minnaert, M.G.J., Houtgast, J., 1966, The Solar Spectrum
2935\,\AA\ to 8770~\AA, National Bureau of Standards, Monograph 61
\ref Perry, C.L., Walter, D.K., Crawford, D.L. 1978, PASP 90, 81
\ref Saxner, M., Hammarb\"ack, G., 1985, A\&A 151, 372
\ref Shu, F.H., Adams, F.C., Lizano, S. 1987, ARAA 25, 23
\ref Thielemann, F.-K. 1993, private communication
\ref Westerlund, B., Garnier, R., Lundgren, K., Pettersson, B., Brey\-sacher,
J.
1988, A\&AS 76, 101 (WGLPB)
\ref Wilson, R.E. 1953, General catalogue of stellar radial velocities,
Mt Wilson observatory papers Vol. VIII, Carnegie Institution, Washington
\endref
\bye